\documentclass[showpacs,aps,preprintnumbers,amsmath,amssymb,twocolumn]{revtex4}
 \usepackage{graphicx}
 \usepackage{dcolumn}
 \usepackage{bm}
 \usepackage{amssymb}
 \usepackage{amsmath}
 \usepackage{amsfonts}
 \usepackage{color}
 \usepackage{natbib}
 \usepackage{url}
 \usepackage{hyperref}
 \usepackage{multirow}
 \usepackage{caption}
 \usepackage{sidecap}
 \usepackage{booktabs}
\usepackage{epstopdf}
\hypersetup{
	colorlinks = true,
	linkcolor=blue,
	urlcolor=blue,
	citecolor=blue,
	}

\newcommand{\refeq}[1]{Eq.(\;\ref{#1})}
\newcommand{\refeqs}[2]{Eqs.\;(\ref{#1}, \ref{#2})}

\newcommand{\refig}[1]{Fig.\;\ref{#1}}

\begin{document}
 \title{PoPe (Projection on Proper elements) for code control:\\ verification, numerical convergence and reduced models.\\
 	Application to plasma turbulence simulations}
 \author{T. Cartier-Michaud}\email{t.cartiermichaud@gmail.com} %\affiliation{CEA/DSM/IRFM, Cadarache, 13108 Saint-Paul-Lez-Durance, France}
 \author{P. Ghendrih} %\affiliation{CEA/DSM/IRFM, Cadarache, 13108 Saint-Paul-Lez-Durance, France}
 \author{Y. Sarazin} %\affiliation{CEA/DSM/IRFM, Cadarache, 13108 Saint-Paul-Lez-Durance, France}
 \author{J. Abiteboul} %\affiliation{CEA/DSM/IRFM, Cadarache, 13108 Saint-Paul-Lez-Durance, France}
 \author{H. Bufferand} %\affiliation{CEA/DSM/IRFM, Cadarache, 13108 Saint-Paul-Lez-Durance, France}
 \author{G. Dif-Pradalier} %\affiliation{CEA/DSM/IRFM, Cadarache, 13108 Saint-Paul-Lez-Durance, France}
 \author{X. Garbet} %\affiliation{CEA/DSM/IRFM, Cadarache, 13108 Saint-Paul-Lez-Durance, France}
 \author{V. Grandgirard} %\affiliation{CEA/DSM/IRFM, Cadarache, 13108 Saint-Paul-Lez-Durance, France}
 \author{G. Latu} %\affiliation{CEA/DSM/IRFM, Cadarache, 13108 Saint-Paul-Lez-Durance, France}
 \author{C. Norscini} %\affiliation{CEA/DSM/IRFM, Cadarache, 13108 Saint-Paul-Lez-Durance, France}
 \author{C. Passeron} %\affiliation{CEA/DSM/IRFM, Cadarache, 13108 Saint-Paul-Lez-Durance, France}
 \author{P. Tamain} %\affiliation{CEA/DSM/IRFM, Cadarache, 13108 Saint-Paul-Lez-Durance, France} 
\affiliation{CEA/DSM/IRFM, Cadarache, 13108 Saint-Paul-Lez-Durance, France} 
\date{\today}
 \begin{abstract}
The Projection on Proper elements (PoPe) is a novel method of code control dedicated to 1) checking the correct implementation of models, 2) determining the convergence of numerical methods and 3) characterizing the residual errors of any given solution at very low cost. The basic idea is to establish a bijection between a simulation and a set of equations that generate it. Recovering equations is direct and relies on a statistical measure of the weight of the various operators. This method can be used in any dimensions and any regime, including chaotic ones. This method also provides a procedure to design reduced models and quantify the ratio costs to benefits. PoPe is applied to a kinetic and a fluid code of plasma turbulence.
\end{abstract}
% Plasma simulation, 52.65.-y
% Plasma turbulence, 52.35.Ra
% chaos numerical simulations, 05.45.Pq
% Numerical methods (mathematics), 02.60.-x
%
\maketitle
%
% % % % % % % % % % % % % % % % % %
%
%%%%%%%%%%
%
%
%\section{Introduction}
%
Numerical simulations are routine tools addressing non-linear problems in most fields of research and industrial applications. Demonstrating their reliability is a major concern \cite{Roache98,Oberkampf2002}. PoPe (Projection on Proper elements) is design to control such numerical experiments.
Code verification, namely checking that the code is solving properly the equations at hand is a straightforward requirement and a critical issue given the complexity of massively parallelized codes. This issue must also encompass numerical convergence and projection of the solution on its validity domain. This method is applied to fusion plasmas where the requirement for precise simulation support is mandatory in view of ITER nuclear experiments \cite{Iked2007}. It is to be underlined that the method is quite general.

PoPe does not require modifying the code nor very specifics runs provided the required data is stored for PoPe post-processing. It can then also be used for in-depth analysis of the simulations, from numerical convergence to determining reduced models. PoPe also provides a series of measures that allows one quantifying the difference between numerical schemes and merits of reduced models. By many aspects it is more appropriate than the present verification standard, the Method of Manufactured Solutions (MMS) \cite{Roache2001,Ricci2015}. In particular, it does not rely on building a target solution with relevant properties, nor does it require modifying the code to enforce these target solutions.

PoPe is represented here using non-linear simulations. As a post-processing option, this versatile tool is shown to be useful to physicists, applied mathematicians and computers scientists. The chosen examples are a pseudo-spectral fluid turbulence code \cite{Sarazin1998} and a more recent semi-Lagrangian kinetic turbulence code \cite{TCM_ESAIM2013}. In the former, the method is presented together with conclusive analysis of the numerical scheme and model reduction capability. In the latter, PoPe is used to verify the code but open issues regarding the numerical convergence are detected and analyzed.\\\\
%
%                  PART 1
%
%**************************************************
\noindent \textbf{PoPe verification of a fluid turbulence code}\\
%**************************************************
%
PoPe is applied to TOKAM2D \cite{Sarazin1998}, a pseudo-spectral code of 2D plasma interchange turbulence, akin to Rayleigh-B\'enard (RB) instability in neutral fluids. It evolves two fields governed by similar equations.
\begin{subequations}
\label{eq:TOKAM}
\begin{eqnarray}
\label{eq:TOKAM1}
\boldsymbol{\partial_t} n  = \alpha_n \boldsymbol{\left\lbrace\phi\right\rbrace}  {n} + D \boldsymbol{\Delta} n  + S_0 \mathbf{S_n} - \sigma_n \mathbf{P_n}~~ \\
\label{eq:TOKAM2}
\boldsymbol{\partial_t} W = \alpha_W\boldsymbol{\left\lbrace \phi\right\rbrace}  {W} + \nu \boldsymbol{\Delta} W -g  \boldsymbol{\partial_y} \mathbf{log}(n) -\sigma_W \mathbf{P_W}~~\\
\label{eq:TOKAM3}
\!\!{W} = \boldsymbol{\Delta} \phi ~;~\boldsymbol{\left\lbrace \phi\right\rbrace  }f = \boldsymbol{\partial_x \phi\partial_y} f - \boldsymbol{\partial_y \phi\partial_x} f~~
\end{eqnarray}
\end{subequations}
Bold characters identify the various operators. This forced turbulence model exhibits an instability that drives the system from a diffusive transport regime, $D$ for particle diffusion and $\nu$ for the viscous damping, to a convective transport regime governed by the Poisson bracket terms \refeq{eq:TOKAM3}. Equations (\ref{eq:TOKAM}) are normalized so that $\alpha_n=\alpha_W =-1$. The $x$-direction runs from the source towards the sink and is that of the effective buoyancy force proportional to $g$. The $y$-direction is periodic.
The plasma model stems from the electron particle \refeq{eq:TOKAM1} and  charge \refeq{eq:TOKAM2} balance equations. The electron density is $n$, $\phi$ is the electric potential, and the vorticity $W$ is related to the polarization density. For RB, $n$ is the temperature field and $\phi$ the stream function. The system is driven out of equilibrium by a source term $\mathbf{S_n}$ constant in $y$, Gaussian in $x$ with fixed normalized width $8.5$. The volumetric non-linear sink terms $\mathbf{P_n}$ and $\mathbf{P_W}$ are specific of the plasma model and account for particle and charge losses in the third direction parallel to the magnetic field, $\mathbf{P_n}= n~\exp(\Lambda-\phi)$ and  $\mathbf{P_W}=1-\exp(\Lambda-\phi)$, $\Lambda$ being a reference value of the electric potential. The values of the various parameters used in the simulation are listed in Appendix \ref{appendix:TOKAM2D control parameters}.
\\\\
%**************************************************
\noindent \textbf{Proper implementation of the model}\\
%**************************************************
%
Let us define ${Z}={(n,W)}$, so that the evolution equation for the $i^{th}$ component of $Z$ takes the general form:
\begin{eqnarray}
%\label{eq:base_TOKAM}
\boldsymbol{\partial_t} Z_i &=& \sum\nolimits_{k} C^{k}_{Z_i} \boldsymbol{\mathcal{O}}_{k}({Z}) 
\label{eq:general_evolution}
\end{eqnarray}
where the operators $\boldsymbol{\mathcal{O}}_{k}({Z})$ are labeled by $k$ and act on $Z$, and where $C^{k}_{Z_i}$ weighs operator $k$ in the evolution equation of $i^{th}$ component of $Z$. Given the ensemble of operators $\left[ \boldsymbol{\mathcal{O}}\right]_{min} = \left[ \boldsymbol{\mathcal{O}}_{k}\right]_{1\le k \le 8}$, the minimal ensemble of operators needed to capture \refeq{eq:TOKAM},  hence $\left[ \boldsymbol{\mathcal{O}}\right]_{min} = \left[ \boldsymbol{\left\lbrace\phi\right\rbrace}n, \boldsymbol{\Delta}n, \mathbf{S_n}, \mathbf{P_n}, \boldsymbol{\left\lbrace\phi\right\rbrace}W, \boldsymbol{\Delta}W, \boldsymbol{\partial_y log}(n), \mathbf{P_W}\right]$, together with the weights :
$\left[C^{k}_{n}\right]_{1\le k \le 4} =\left[\alpha_n, D, S_0, -\sigma_n\right]$, $\left[C^{k}_{n}\right]_{5\le k \le 8} =\left[0, 0, 0, 0\right]$, and:  $\left[C^{k}_{W}\right]_{1\le k \le 4} =\left[0, 0, 0, 0\right]$,  $\left[C^{k}_{W}\right]_{5\le k \le 8} =\left[\alpha_W, \nu, -g, -\sigma_W\right]$.
For a code solving the generic equation \refeq{eq:general_evolution}, PoPe only requires saving within the code the computed values of the field ${Z}$ and its time derivative $\boldsymbol{\delta_t} Z$ (readily obtain with finite differences) at synchronized locations and times. This must be done in such a way that, using this output data, an ensemble of operators $[\boldsymbol{\mathcal{O}}]_{N_O}$ acting on ${Z}$ can be computed independently. The latter calculation can readily be performed with fully verified and controlled schemes. The projection of the numerical values of $\boldsymbol{\delta_t} Z$ distributed in space and time ($N_s\! \times\! N_t$ points) on an ensemble of numerically estimated operators $[\boldsymbol{\mathcal{O}}]_{N_O}$, which includes $[\boldsymbol{\mathcal{O}}_{min}]$, then yields the sets $[p_{Z_i}^{k}(j)]$ where  $p_{Z_i}^{k}(j)$ is the $j^{th}$ projection with $j$ ranging typically from $1$ to $N_t\! \times\! N_s / N_O \gg 1$. %
This projection is performed by solving linear systems with the $p_{Z_i}^{k}(j)$ considered as unknown and $\boldsymbol{\delta_t} Z_i$ evaluated on the $j^{th}$ subset of points as right hand side. Since the weights of the various operators $\boldsymbol{\mathcal{O}}_k$ in \refeq{eq:general_evolution} are constant and equal to $C_{Z_i}^{k}$, and since the model equations must be verified at all times and all locations in space, the statistical data set $[p_{Z_i}^{k}(j)]$, is the Probability Density Function of numerical realizations of $C_{Z_i}^{k}$.
Of immediate relevance are then the mean-value of the computed coefficients $c_{Z_i}^k= \left\langle p_{Z_i}^{k}(j)\right\rangle_j $, the characteristic error $\Delta C_{Z_i}^{k}= | C_{Z_i}^{k}-c_{Z_i}^{k} |  $ and the data scatter evaluated by $\delta c_{Z_i}^{k}=\left\langle || c_{Z_i}^{k}-p_{Z_i}^{k}(j) || \right\rangle_j $.
The verification with PoPe of the proper implementation of the operator $\boldsymbol{\mathcal{O}}_k$ in the evolution equation of $Z_i$ is then characterized by $\Delta C_{Z_i}^{k} / C_{Z_i}^{k} \ll 1$ and $\delta c_{Z_i}^{k} / c_{Z_i}^{k} \ll 1$. Furthermore, these values quantify the error introduced when stepping from the equations to their numerical integration.
\begin{table}[h]
	\hspace{0.mm}
	\begin{tabular}{c|c|c|c|c|}
		%	\hline
		%		\hline
		& $\boldsymbol{\mathcal{O}}_k$&  \refeq{eq:TOKAM1} run &  Error: & Fluctuation \\
		k & operator & value: $C_n^{k}$&  $\Delta C_n^{k}$ &  level: $\delta c_n^{k}$ \\
		\hline & &   &  	&\\ 
		%		\hline	
		1 & $\boldsymbol{\left\lbrace \phi\right\rbrace  n} $ 	& $-1$ &  $ 1.99 ~ 10^{-9} $& $ 3.01~10^{-9}$\\
		%		\hline
		2 & $\boldsymbol{\Delta n}$			& $0.02$ & $2.23~ 10^{-11}$ &$4.46~10^{-11}$\\
		%		\hline
		3 & $\boldsymbol{S_n}$     		& $0.01$ &  $5.86 ~ 10^{-11}$ &$1.22~10^{-10}$\\
		%		\hline
		4 & $\boldsymbol{P_n}$		& $-1.13 ~ 10^{-5}$& $6.51~ 10^{-14}$ & $1.33~10^{-13}$\\
		%		\hline
		5 & $\boldsymbol{\left\lbrace \phi\right\rbrace  W}$& $0$ &$3.06 ~10^{-7}$ &$7.64 ~10^{-7}$\\ 
		%	\hline
		6 & $\boldsymbol{\Delta W}$			& $0$ & $1.09~10^{-8}$ &$3.21~10^{-8}$\\
		%	\hline
		7 & $\boldsymbol{\partial_y Log(n)}$			& $0$ & $1.72~10^{-11}$ &$7.02 ~10^{-10}$\\
		%	\hline
		8 & $\boldsymbol{P_W}$			& $0$ & $1.03~10^{-11}$ &$3.95~10^{-11}$\\
		%	\hline
	\end{tabular}
	\caption{Verification of TOKAM2D,  $N_x = N_y = 256$, $N_x/L_x=N_y /L_y =2$ and $\Delta t = 2^{-5}$}
	\label{tab:verification TOKAM2D}
	\vspace{-15pt}
\end{table}
TOKAM2D is verified with a run of the code performed with a small time step $\Delta t = 2^{-5}$, while the value used in production runs is $\Delta t = 1$, and a mesh comparable to a standard production mesh $N_x \!\times\! N_y = 2^8 \!\times\! 2^8$ for a box size $L_x \!\times\! L_y = 2^7 \!\times\! 2^7$. For this simulation, the most unstable mode $k_y$ is such that $N_y /(k_y L_y) \approx 13.6$ and its growth rate is $\gamma_{k_y} \approx 7. ~10^{-4} \ll 1$. The verification of the numerical implementation of \refeq{eq:TOKAM1} are summarized in table \ref{tab:verification TOKAM2D}. One finds that for each operator the error $\Delta C_n^k$ is small as well as the fluctuation level $\delta c_n^k$. Furthermore, one finds that $\delta c_n^k > \Delta C_n^k$ which means that the exact value stands in the region of maximum probability of the PDF. It is to be noted that \refeqs{eq:TOKAM1}{eq:TOKAM2} are coupled in the evolution scheme that is based on the analytical calculation of the eigenvectors and eigenvalues of the linearized equation. 
However, the computed solution of \refeq{eq:TOKAM1} exhibits a very small dependence on the operators  $\boldsymbol{\mathcal{O}}_{5 \le k\le 8}$ that govern the evolution of $W$ since their computed weight is close to zero as required. Pope thus allows one to recover with good precision the weight of the various operators of the evolution equation, the implementation of the equations is thus correct.

When analyzing the PDF of the error of the coefficients of \refeq{eq:TOKAM1}, one finds a Poisson-like distribution for the Poisson bracket $\boldsymbol{\left\lbrace\phi\right\rbrace}  {n}$ with $\Delta C_n^1 \ge 0$, compatible with a rare deviation from the exact value, while the error for the coefficient $\boldsymbol{\left\lbrace\phi\right\rbrace}  {W}$ is more Gaussian like, which can be interpreted as a random value in the vicinity of zero. Such marked differences in the PDFs remain to be fully analyzed and understood.
\begin{figure}[h]
	\vspace{-10pt}  
	\resizebox{0.4975\textwidth}{!}{%
	\includegraphics{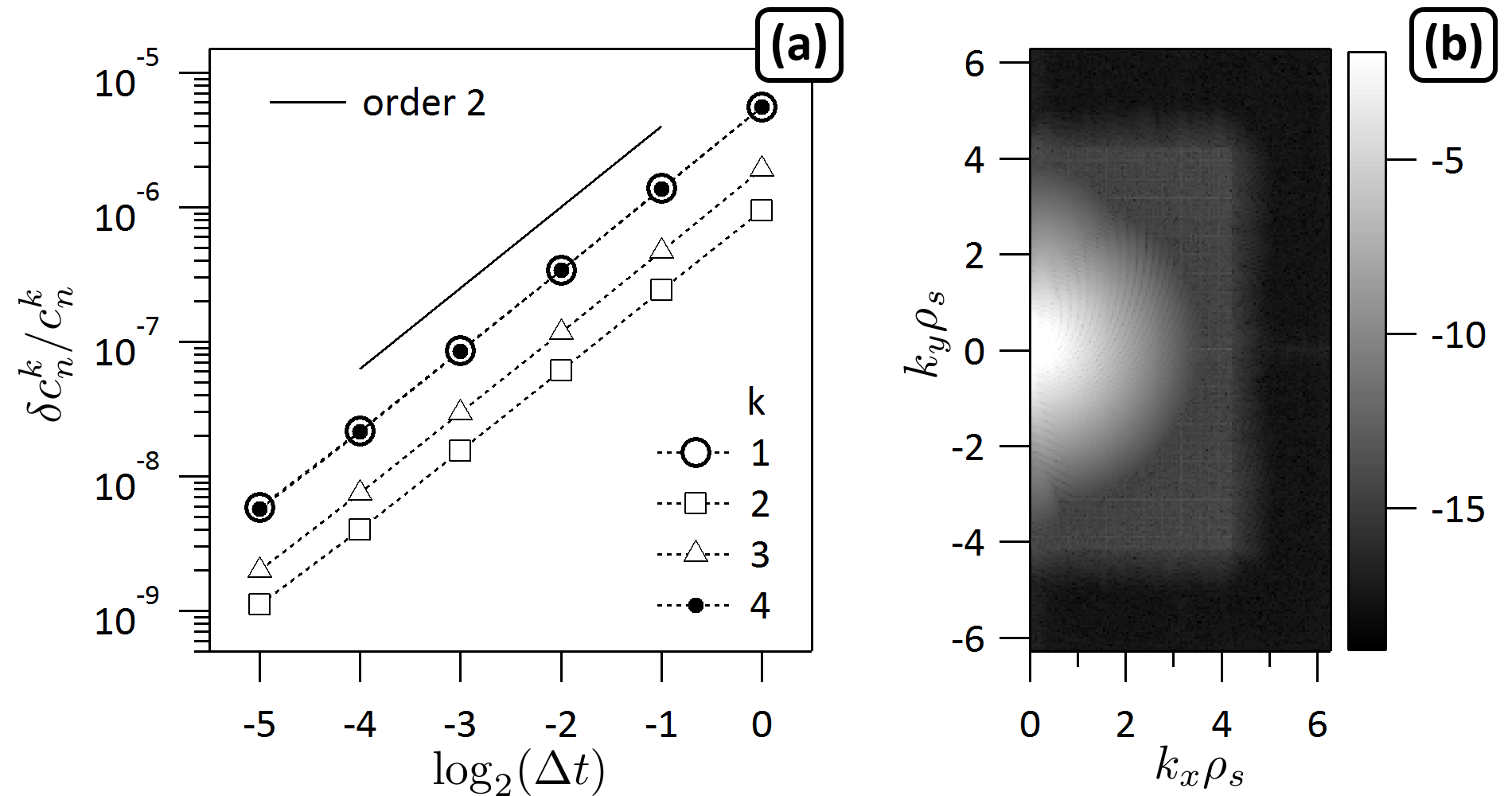}}
	\caption{(a) relative error on TOKAM2D coefficients in \refeq{eq:TOKAM1} when scanning the time step $\Delta t$. The order 2 time integration scheme is verified. (b) 2D Fourier spectrum of $\sum\nolimits_{k} C^{k}_n \boldsymbol{\mathcal{O}}_{k}({Z})$ on a $256\times256$ mesh.}
	\label{fig:error and 2D spectrum}
	\vspace{-5pt}  
\end{figure}

%
%                  PART 2
%
%**************************************************
\noindent \textbf{Accuracy of the numerical scheme}\\
%**************************************************
%
Given the code verification, PoPe then provides a means to evaluate the accuracy of the numerical scheme. This is illustrated by computing the error $\Delta C_n^{1\le k \le 4}$ and the fluctuation level $\delta c_n^{1\le k \le 4}$, when scanning the time step $\Delta t$. The numerical evolution scheme is an order 2 predictor-corrector. It is found to be properly implemented since one finds that the relative error $\delta c_n^k / c_n^k$ scales like $\Delta t^2$, see \refig{fig:error and 2D spectrum}(a). The characteristic error $\Delta C_n^{k}$ exhibits the same dependence as the fluctuation level $\delta c_n^{k}$, which means that the improved matching is not governed by a change in the error statistics. Even if the $C_n^k$ exhibits differences of several orders of magnitude, relative errors are comparable. 

\begin{figure}[h]
	\vspace{-20pt}
	\resizebox{0.4975\textwidth}{!}{%
		\includegraphics{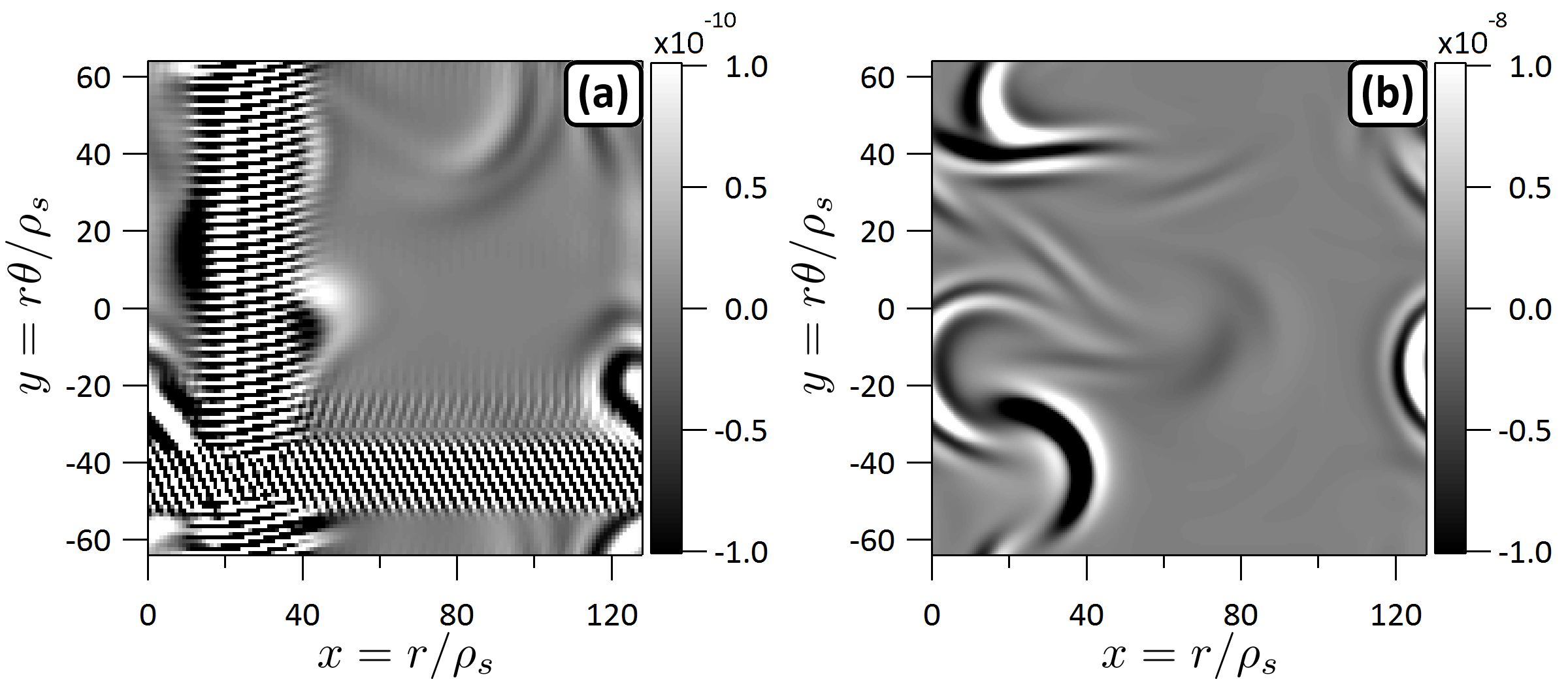}}
	\caption{Residual error $\varepsilon_n$ \refeq{eq:residual error} for two simulations, (a) $128\times128$, (b) enhanced resolution, $256\times256$.}
	\label{fig:residual error real space}      
	\vspace{-10pt}
\end{figure}

When analyzing the projection on operators that are absent from \refeq{eq:TOKAM1}, 
% $\Delta C_n^{5\le k \le 8}$,
$\Delta C_n^k$ and $\delta c_n^k$ scale like $\Delta t^2$ but the criterion $\delta c_n^k / |c_n^k| = \delta c_n^k / \Delta C_n^k $ does not. Indeed, $\delta c_n^k /  c_n^k \ll 1$ identifies operators used in the equation while $\delta c_n^k / \Delta C_n^k \gtrsim 1$ identifies absent operators for which the error in determining zero and its fluctuation level are comparable.
\\\\
%
%                  PART 3
%
%**************************************************
\noindent \textbf{Analysis of the residual error}\\
%**************************************************
Model and numerical scheme implementation verification is only complete if and only if the residual error, namely $\varepsilon_i$, is small. This residual error can also be analyzed for in depth control of the verification procedure, it is defined such as:
\begin{eqnarray}
\label{eq:base_TOKAM}
\varepsilon_i = \boldsymbol{\delta_t} Z_i -\! \sum\nolimits_{k} C^{k}_{Z_i} \boldsymbol{\mathcal{O}}_{k}({Z}) = \mathcal{P} \varepsilon_i + \left( 1-\mathcal{P}\right)  \varepsilon_i
\label{eq:residual error}
\end{eqnarray}
given $\mathcal{P}$ the projection operator on the subspace where the solution is reliable. The residual error thus addresses the error made by TOKAM2D on the time derivative $ \boldsymbol{\delta_t} Z_i$. It highlights fast displacement regions with sharp changes of the magnitude of $Z_i$, hence fronts. The residual error is thus a measure of the error made on the front location and height, \refig{fig:residual error real space}.  \\

\begin{figure}[h]
	\vspace{-5pt}
	\resizebox{0.4975\textwidth}{!}{%
	\includegraphics{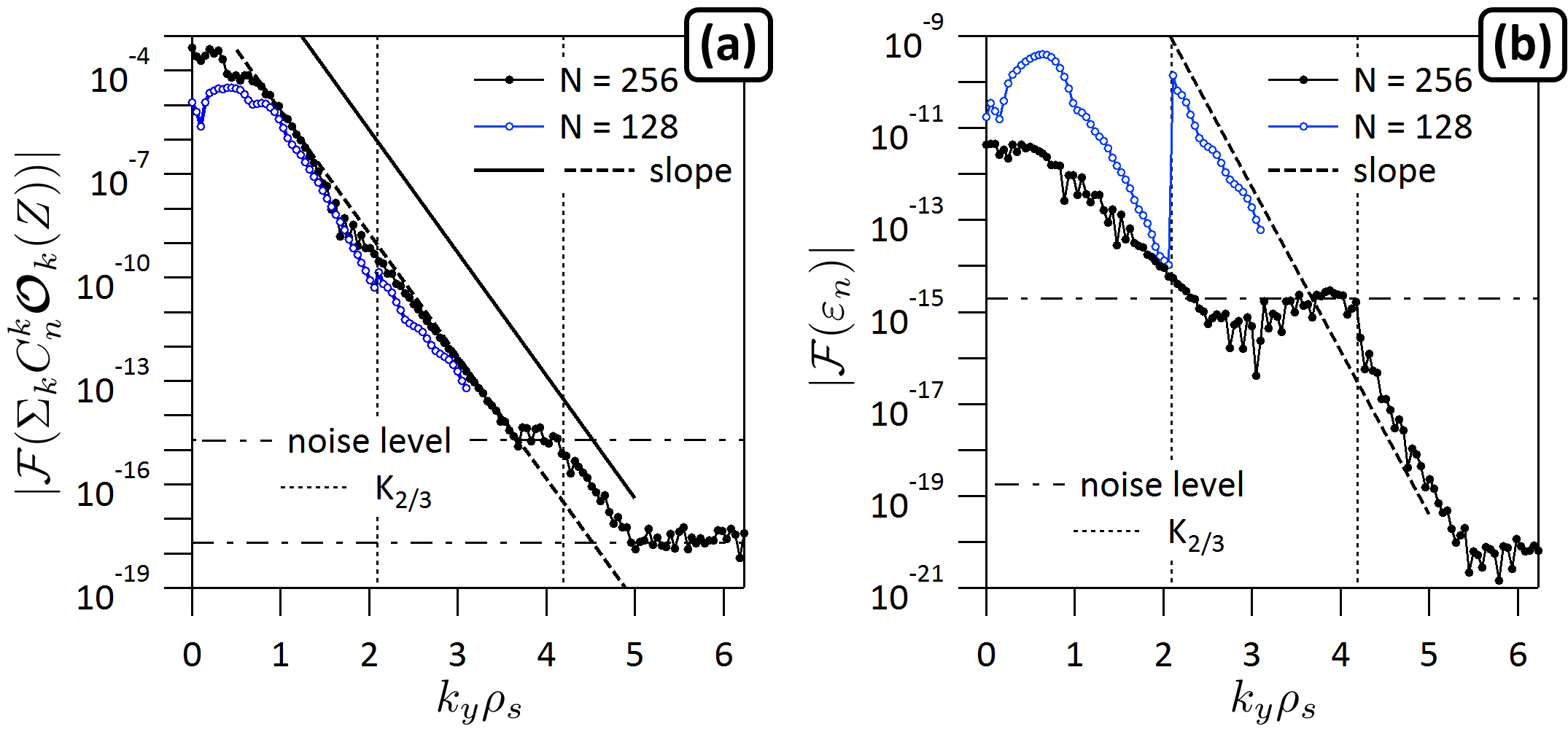}}
	\caption{(a) Fourier transform of $\sum\nolimits_{k} C^{k}_n \boldsymbol{\mathcal{O}}_{k}({Z})$, RHS of \refeq{eq:TOKAM1}. (b): Fourier transform of residual error $\varepsilon_n = \boldsymbol{\delta_t} n -\! \sum\nolimits_{k} C^{k}_n \boldsymbol{\mathcal{O}}_{k}({Z})$, RHS of \refeq{eq:residual error}.}
	\label{fig:spectre2D}      
	\vspace{-5pt}
\end{figure}

This projection operator $\mathcal{P}$ retaining the reliable part of the solution is readily determined in pseudo-spectral codes like TOKAM2D since the de-aliasing rule sets 1/3 of the spectra at zero and is applied at each calculation of the non-linear terms. This is noticeable in the 2D space spectrum of $\sum\nolimits_{k} C^{k}_i \boldsymbol{\mathcal{O}}_{k}({Z})$, %$\boldsymbol{\partial_t} Z_i$,
$|\mathcal{F}(\sum\nolimits_{k} C^{k}_i \boldsymbol{\mathcal{O}}_{k}({Z}))|$, with the square shaped boundary to the region at machine precision noise \refig{fig:error and 2D spectrum}(b). Prior to this region, one can notice a flat shoulder at the numerical noise level of the scheme ($\approx 10^{-15}$) for $k$-values below the cut-off. In this properly discretized example, the ill-computed output data belongs to the region where numerical noise is reached. This feature is readily seen on \refig{fig:spectre2D}. One finds that the spectra are quite comparable for different step size in space (a), while the residual error exhibits a marked difference in shape (b). For the low resolution case, $N=128$, the residual error exhibits a marked discontinuity at the cut-off value of the wave vector which shows that the projection $\mathcal{P}$ is removing a meaningful part of the spectrum. Conversely, for $N=256$, the residual error exhibits a smooth decay from the numerical to the machine noise levels (b).

In real space the residual error $\varepsilon_n$, \refig{fig:residual error real space}(a), allows one identifying accuracy issues, hence for the low resolution case, $N=128$, aliasing patterns indicate that space resolution improvement is required. Evidence of a proper discretization is given by the residual error for $N=256$, \refig{fig:residual error real space}(b). The residual error then exhibits a dipolar structure suggesting that the front location displacement is underestimated while its shape is properly captured. Further numerical scheme improvement is then governed by more accurate time stepping.\\\\
%
% PART 4
%
%**************************************************
\noindent \textbf{PoPe verification of the kinetic code TERESA}\\
%**************************************************
We consider here a $3D$ kinetic code, $2D$ in space, the angle $\xi$ and the radial label $\psi$, and $1D$ in energy $E$ for the various classes of particles. By many aspects, the physical model addressed by the TERESA code can be understood as a kinetic version of the Rayleigh-B\'enard problem. Regarding plasma physics, it is a low frequency model of turbulence for deeply trapped particles in the harmonic oscillator regime of the tokamak confinement geometry. The equations implemented in the TERESA code are the Vlasov equation for the particle distribution function and a Poisson equation yielding the electrostatic potential.
\begin{subequations}
	\label{eq: TERESA}
\begin{eqnarray}
\boldsymbol{\partial_t} \bar{f} = -\left( \boldsymbol{\partial_\psi}\bar{\boldsymbol{\phi}} \boldsymbol{\partial_\xi} \bar{f}  - \boldsymbol{\partial_\xi}\bar{\boldsymbol{\phi}} \boldsymbol{\partial_\psi} \bar{f} + (\omega_D E) \boldsymbol{\partial_\xi} \bar{f}\right) 
\label{eq:base_TERESA_VLASOV} \\
\int\!\! \boldsymbol{\mathcal{J}}\bar{f} dE -1 =- \left( C_e \left(\phi -  \lambda \left\langle \phi \right\rangle_\xi  \right)  - C_i \boldsymbol{\Delta_{J}} \phi \right) 
\label{eq:base_TERESA_quasineutrality_mean}
\end{eqnarray}
\end{subequations}
where the bar quantities in \refeq{eq:base_TERESA_VLASOV} are obtained from the no bar quantities by a double average on high frequency contributions to particle motion. For any given function U depending on space coordinates, it is computed as:
\begin{subequations}
\begin{eqnarray}
	\label{def: bar operator}
	\bar{U}\!\!&=&\!\!\boldsymbol{\mathcal{J}}U = \left( 1-\frac{E\delta_0^2}{4} \boldsymbol{\partial^2_\psi} \right)^{-1}\!\! \left(1-\frac{E\rho_0^2}{4} \boldsymbol{\partial^2_\xi} \right)^{-1}\!\!\!\! \!U 
	\\
	\boldsymbol{\Delta_{J}} \!\!&=&\!\! \rho_0^2\boldsymbol{\partial_\xi^2} + \delta_0^2\boldsymbol{\partial_\psi^2}
	\label{def: Delta}
\end{eqnarray}
\end{subequations}
The operator $\boldsymbol{\Delta_{J}}$ that appears in \refeq{eq:base_TERESA_quasineutrality_mean}, defined in \refeq{def: Delta}, is related to the polarization of the particle trajectories during the high frequency quasi-periodic motion. Its expression is thus closely related to the averaging operator \refeq{def: bar operator}. 

\begin{SCfigure}[][h]
\centering
\includegraphics[width=.25\textwidth]{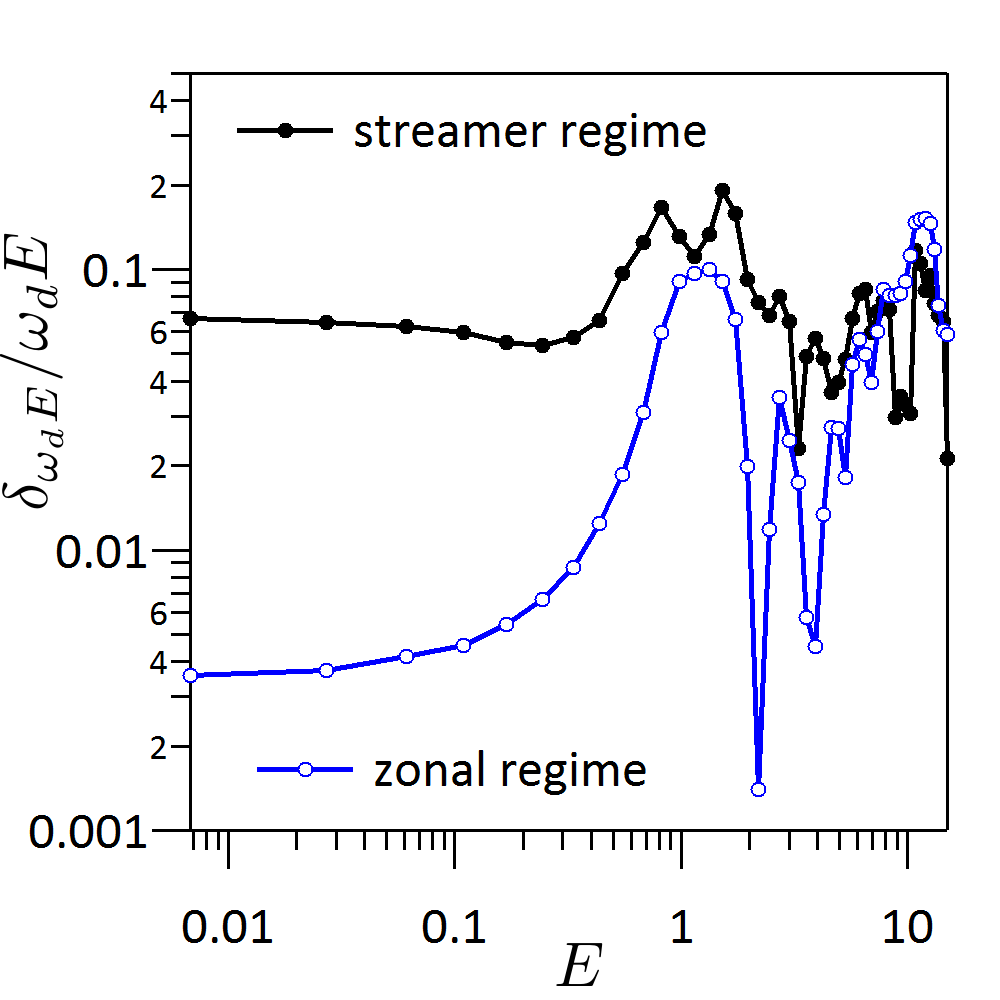}
\caption{Relative error of the weight $\omega_d E$ in \refeq{eq:base_TERESA_VLASOV} with respect to the energy dimension.} \label{fig:vlasov}
\vspace{-5pt}
\end{SCfigure}

Compare to TOKAM2D, TERESA verification is addressed here by retaining only the dependence of the relative error on the energy. 
The relative error of the weights of the Vlasov equation, $1$, $-1$ for the Poisson bracket and $\omega_D~E$ for the drift motion, ranges from $10^{-1}$ to $10^{-5}$. This indicates that the model is correctly implemented. The energy dependence of the error is two-fold. Regarding the Poisson bracket terms, one finds that improving the precision allows one reducing the error. Conversely, for the $\xi$-convection, weight $\omega_D E$, the error remains significant with an increase towards the small values of $E$ for solution with large convection cells (streamer solutions), see \refig{fig:vlasov}. This limitation is understood given the quasi-steady state solution of the TERESA model with large convection cells that generates fronts in the distribution function at the mesh resolution. These fronts are approximately aligned along constant $\xi$-lines and are convected in $\xi$. Such a behavior was not anticipated and the interpolation scheme that is presently used in the TERESA semi-Lagrangian code is not suited for such sharp front solutions. Small errors in the location of the step in the distribution value thus generates a large error that cannot be reduced since the width of the front is the mesh resolution.
\vspace{-10pt}

%
% PART V
%
%**************************************************
\noindent \textbf{Quantifying reduced models with PoPe}\\
%**************************************************
%
The relation established between the code output and the weight of the various operators can be used to evaluate reduced models rather than the source model as done in the verification step. 
Building a suit of models from first principles to operation developed in a consistent fashion will be mandatory for ITER, and even more in DEMO fusion reactors with limited diagnostics. In such a framework, one presents here the best fitting Lotka-Volterra model \cite{Lotka10,Volterra26} of TOKAM2D intermittent regime.

\begin{figure}[h]
	\vspace{-5pt}	
	\resizebox{0.4975\textwidth}{!}{%
		\includegraphics{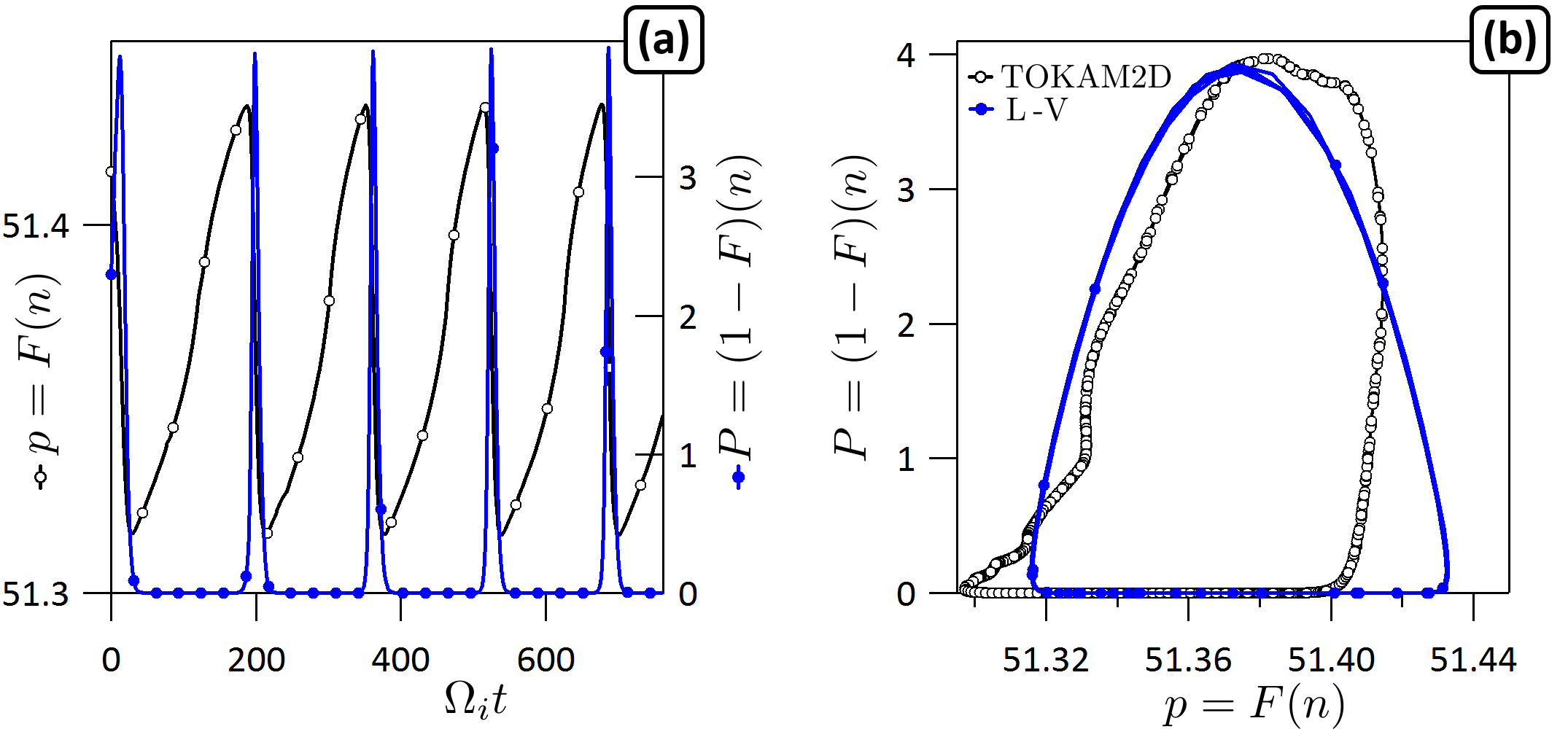}}
	\caption{(a) time trace of predator, $P=(1-F)n$ closed symbols, prey, $p=F(n)$ open symbols. (b) Predator versus prey cycle obtained with TOKAM2D, open symbols, Lotka-Volterra reduced model determined by PoPe, closed symbols.}
	\label{figure_PP}      
	\vspace{-5.5pt}
\end{figure}

We consider TOKAM2D run in a regime with a large viscosity, and therefore with a reduced number of degree of freedom. A decrease of turbulent activity and chaos is expected \cite{Maschke1982}. In the chosen example, an intermittent behavior with periods of quasi-steady convective transport followed by rapid bursts of turbulence is observed. This governs a quasi-periodic cycle of relaxation events reminiscent of the predator prey dynamics \cite{Diamond1995}. To test the validity of this description one determines with PoPe the best set of weights of the Lotka-Volterra predator-prey model to capture this behavior. For this analysis, the code data is reduced to two fields: the prey $p$ which is the energy of the set of harmonics of the most unstable mode of the Fourier transform of the density \refeq{eq:TOKAM1}, $p=F(n)$, and the predator $P$ standing for the energy of all the other Fourier modes, $P=(1-F)(n)$, see Fig. (\ref{figure_PP}). From this figure one can notice that minimizing the residual error yields a Lotka-Volterra cycle that closely matches the TOKAM2D cycle. Similarly, alternative predator-prey models can then be tested.
\\\\
%
% PART 6
%
%%%%%%%%%%
\noindent \textbf{Summary}\\
%%%%%%%%%%
It is possible to numerically recover equations from a set of data, in the present case, generated by simulations. If one knows the equations to be implemented in the code, PoPe is a powerful method to monitor the accuracy of the numerical methods by studying the weights of the various operators and recovering the values that are used as simulation input. Such direct data-mining method is intrinsic conversely to benchmark verification that relies on external verification.
Using the same procedure, PoPe provides a method to determine reduced models that capture a fraction of the dynamics observed in the data.
From a computational point of view, this method is generic, straight forward to implement and with an overall low computational cost.
Regarding plasma turbulence simulations, we have shown that a code can be verified regarding the equation implementation despite using an unoptimized numerical scheme as in the TERESA case. For the other example, TOKAM2D, good control of the simulation is achieved. This leads us to understand the results in terms of statistical distribution of the control parameters in both original and reduced models. By quantifying these statistics PoPe provide a means to adapt the simulation accuracy to a target precision.
\clearpage

\acknowledgements
This work was granted access to the HPC resources of Aix-Marseille Université supported by the project Equip@Meso (ANR-10-EQPX-29-01) of the program "Investissements d'Avenir" supervised by the Agence Nationale pour la Recherche. The authors acknowledge fruitful interactions at La Maison de la Simulation (Saclay, France), the Festival de Th\'eorie (Aix-en-Provence, France) and financial support of the A*MIDEX project "Kinetic Fluid Computing and signature in measurements" (ANR-11-IDEX-0001-02) funded by "Investissements d'Avenir".

\bibliographystyle{unsrt}
\bibliography{biblio_PoPe}

\appendix

\section{TOKAM2D parameters}
\label{appendix:TOKAM2D control parameters}
\begin{table}[h]
	\vspace{-5.mm}
	\begin{tabular}{c|c|c|c|}
		%	\hline
		%		\hline
		\refeq{eq:TOKAM1} &  value & \refeq{eq:TOKAM2} &  value  \\
		\hline &   &  	&\\ 
		%		\hline	
		$\alpha_n $ 	& $-1$ &  $ \alpha_W $&$-1$\\
		%		\hline
		$D$			& $0.02$ & $\nu$ & $0.01$\\
		%		\hline
		$S_0$     		& $0.01$ &  $g$ &$  5.723 ~ 10^{-4}$\\
		%		\hline
		$\sigma_n$		& $1.135 ~ 10^{-5}$& $\sigma_W$ &$ - 1.135 ~ 10^{-5}$\\
		%		\hline
		%	\hline
	\end{tabular}
%	\caption{Input parameters of TOKAM2D}
	\label{tab:control_parameters}
\end{table}
	
\end{document}